# Multigrid Monte Carlo with higher cycles in the Sine Gordon model

Martin Grabenstein and Bernhard Mikeska

*II. Institut für Theoretische Physik, Universität Hamburg,*

*Luruper Chaussee 149, W-2000 Hamburg 50, Germany*

## Abstract

We study the dynamical critical behavior of multigrid Monte Carlo for the two dimensional Sine Gordon model on lattices up to $128 \times 128$. Using piecewise constant interpolation, we perform a W-cycle ($\gamma = 2$). We examine whether one can reduce critical slowing down caused by decreasing acceptance rates on large blocks by doing more work on coarser lattices. To this end, we choose a higher cycle with $\gamma = 4$. The results clearly demonstrate that critical slowing down is not reduced in either case.

Typeset using REVTeX

Computer simulations with local algorithms in statistical mechanics and lattice gauge theory close to a critial point suffer from critical slowing down (CSD). Besides other nonlocal methods, multigrid Monte Carlo algorithms were introduced to overcome this problem [1].

An attempt to understand *why* multigrid Monte Carlo is successful in beating CSD for some models while it does not work as well for others, was made in [2]: An analytic calculation and analysis of acceptance rates for nonlocal Metropolis updating was performed. It was argued that for a critical model with Hamiltonian $\mathcal{H}(\phi)$, CSD will occur if the expansion of $\langle\mathcal{H}(\phi+\psi)\rangle$ in terms of the shift $\psi$ contains a relevant (mass) term. If such a term is present, Metropolis step sizes $\varepsilon(L_B)$ on block lattices with increasing block size $L_B$ have to be scaled down like $\varepsilon(L_B) \sim L_B^{-1}$ in order to obtain block size independent acceptance rates. This strong decrease of step sizes on large blocks was found in several models for smooth and piecewise constant interpolation. One of these models is the Sine Gordon model in two dimensions.

The purpose of this note ist twofold: First, we want to check the prediction [2] that a W-cycle (cycle control parameter $\gamma = 2$) with piecewise constant interpolation will not eliminate CSD in the rough (massless) phase of the Sine Gordon model. (This is a method that eliminates CSD in the Gaussian model.) Secondly, we want to ask the question whether one can circumvent slowing down caused by too small steps on large blocks by accumulating many of these steps randomly. If this would be possible, the acceptance problem could be solved by doing more work on coarser lattices.

A constant accumulated step size on all length scales can be achieved in the following way: For step sizes scaling down like $\varepsilon(L_B) \sim L_B^{-1}$ and a coarsening by a factor of two, the Metropolis step size on a next coarser grid is too small by a factor of two. Assuming a random-walk like accumulation of the steps, one can expect to compensate for this decrease by increasing the number of updates on the next coarser grid by a factor of four. This can be achieved by a higher cycle with cycle control parameter $\gamma = 4$. The rule for higher cycles is that from an intermediate block lattice one proceeds $\gamma$ times to the next coarser lattice before going back to the next finer lattice. In this way $\gamma$ times more updates on each coarser lattice are performed.

For a recursive multigrid algorithm, the computational effort is $\sim L^d$ for $\gamma < 2^d$ and $\sim L^d \log L$ for $\gamma = 2^d$ in $d$ dimensions [3]. Therefore a higher cycle with $\gamma = 4$ is practical for $d > 2$ and borderline practical for $d = 2$.

The 2-d Sine Gordon model is defined on an $L \times L$ lattice $\Lambda_0$ via the partition function

$$Z = \int \prod_{x \in \Lambda_0} d\phi_x \exp(-\mathcal{H}(\phi)) \;, \qquad (1)$$

with the Hamiltonian

$$\mathcal{H}(\phi) = \frac{1}{2\beta} \sum_{\langle x,y \rangle} (\phi_x - \phi_y)^2 - \zeta \sum_x \cos 2\pi \phi_x \;. \qquad (2)$$

The first sum runs over all bonds in the lattice. From the point of view of statistical mechanics, this system can be considered as a 2-d surface in a periodic potential. The model exhibits a Kosterlitz-Thouless phase transition at $\beta_c(\zeta)$. In the limit of vanishing fugacity $\zeta$, $\beta_c$ takes the value $2/\pi = 0.6366\ldots$. For $\beta > \beta_c$ the model is in the rough (massless) phase.

There, the cosine-term of the Hamiltonian is irrelevant in the renormalization group sense. The system is critical and has the same long distance behavior as the massless Gaussian model. The fluctuations of the surface are given by the surface thickness

$$\sigma^2 = \left\langle (\phi_x - \overline{\phi})^2 \right\rangle , \qquad (3)$$

where $\overline{\phi}$ denotes the average of the field over the lattice. In the rough phase, $\sigma^2$ scales with $\log L$ [4].

Our simulations are organized as follows: In order to allow for high cycle control parameters $\gamma$, we use a recursive multigrid algorithm, piecewise constant interpolation and a staggered coarsening with a factor of two as described in [3]. As pre-smoothing and post-smoothing operation, we choose a sweep of single hit Metropolis updates. The maximum Metropolis step size $\varepsilon(L_B)$ is scaled down like $L_B^{-1}$. Then, acceptance rates of approximately 50% are observed on all block lattices, in accordance with the theoretical analysis of [2].

We study the dynamical critical behavior of the different versions of the algorithm in the rough phase, where the correlation length is infinite and the physical length scale is set by the linear size of the lattice $L$. Thus, we expect the autocorrelation time $\tau$ to diverge with the dynamical critical exponent $z$ like $\tau \sim L^z$.

The simulations were performed at $\beta = 1.0$, $\zeta = 0.5$. This is deep in the rough phase. Note that in the limit $\zeta \to \infty$ which corresponds to the discrete Gaussian model [5], the critical coupling is $\beta_c = 0.7524(8)$ [6]. Starting from an ordered configuration, measurements were taken after equilibration at each visit of the finest lattice. From the autocorrelation function for the observable $A$

$$\rho_A(t) = \frac{\langle A_s A_{s+t} \rangle - \langle A \rangle^2}{\langle A^2 \rangle - \langle A \rangle^2} , \qquad (4)$$

we computed the integrated autocorrelation times

$$\tau_{int,A} = \frac{1}{2} + \sum_{t=1}^{\infty} \rho_A(t) \qquad (5)$$

and the corresponding errors by a window method [7] with a self-consistent truncation window of $4\tau_{int}$ for the energy $E = L^{-2} \sum_{\langle x,y \rangle} (\phi_x - \phi_y)^2$ and the surface thickness $\sigma^2$. We checked that the autocorrelation functions for $\sigma^2$ showed an exponential decay.

The numerical results are given in Table I for $\gamma = 2$ and in Table II for $\gamma = 4$. The autocorrelation times $\tau_{int}$ are measured in the number of visits on the finest lattice. Note that our runs are longer than $10\,000$ $\tau_{int}$ (longer than $4\,000$ $\tau_{int}$ on the $128^2$ lattice). Fig.1 shows the dependence of $\tau_{int,\sigma^2}$ on $L$ for $\gamma = 2$ and $\gamma = 4$. For comparison, we plotted lines which correspond to $z = 2$.

If we fit our data for the autocorrelation time of the surface thickness in the range $32 \leq L \leq 128$ with the Ansatz $\tau_{int,\sigma^2} = cL^z$, we obtain $z = 1.86(4)$ with $\chi^2/\text{dof} = 0.22$ for the W-cycle ($\gamma = 2$), and $z = 1.86(4)$ with $\chi^2/\text{dof} = 4.6$ for the higher cycle ($\gamma = 4$). The uncertainty in $z$ is dominated by the relative error of $\tau_{int}$ on the largest lattice, which is about 6% in both cases. We therefore estimate

$$z = 1.9(1) \quad \text{for the W-cycle } (\gamma = 2) ,$$
$$z = 1.9(1) \quad \text{for the higher cycle } (\gamma = 4) .$$

Thus, as already predicted in [2], CSD in the rough phase of the Sine Gordon model is not reduced by a W-cycle with piecewise constant interpolation. According to the acceptance analysis, we would expect the same result with smooth interpolation. Moreover, the results clearly show that compensating for decreasing acceptance rates on large blocks by choosing a higher cycle with $\gamma = 4$ does not improve the dynamical critical behavior of the algorithm. We conclude that a random-walk like argumentation as stated above is not correct in the case of the Sine Gordon model.

## ACKNOWLEDGMENTS

We would like to thank G. Mack, S. Meyer and K. Pinn for helpful discussions. M. G. would like to thank A. Brandt for the kind hospitality and the stimulating atmosphere at the Weizmann Institute. B. M. is grateful to D. W. Heermann for his encouragment and discussions during the early stage of this study. This work was supported in parts by the Deutsche Forschungsgemeinschaft, the German Israeli Foundation and the German Scholarship Foundation. Our simulations were performed on hp 9000/730 RISC workstations at DESY.

TABLES

TABLE I. Numerical results for the W-cycle ($\gamma = 2$) in the 2-d Sine Gordon model on $L \times L$ lattices in the rough phase, $\beta = 1.0$, $\zeta = 0.5$.

| $L$ | statistics | discarded | $E$ | $\tau_{int,E}$ | $\sigma^2$ | $\tau_{int,\sigma^2}$ |
|---|---|---|---|---|---|---|
| 4 | 25 000 | 2 000 | 0.934(4) | 0.90(3) | 0.268(1) | 0.96(3) |
| 8 | 50 000 | 2 000 | 0.986(1) | 0.97(2) | 0.3809(9) | 1.35(3) |
| 16 | 100 000 | 2 000 | 0.9956(5) | 1.04(2) | 0.4896(7) | 2.70(8) |
| 32 | 300 000 | 2 000 | 0.9987(2) | 1.03(1) | 0.5996(7) | 8.54(19) |
| 64 | 500 000 | 2 000 | 0.99945(5) | 1.04(1) | 0.7105(10) | 30.5(1.0) |
| 128 | 500 000 | 4 000 | 0.99966(3) | 1.04(1) | 0.8218(19) | 113.7(6.9) |

TABLE II. Numerical results for the higher cycle with $\gamma = 4$ in the 2-d Sine Gordon model on $L \times L$ lattices in the rough phase, $\beta = 1.0$, $\zeta = 0.5$.

| $L$ | statistics | discarded | $E$ | $\tau_{int,E}$ | $\sigma^2$ | $\tau_{int,\sigma^2}$ |
|---|---|---|---|---|---|---|
| 4 | 25 000 | 2 000 | 0.940(4) | 0.89(3) | 0.268(1) | 0.91(3) |
| 8 | 25 000 | 2 000 | 0.986(2) | 0.94(3) | 0.380(1) | 1.14(4) |
| 16 | 25 000 | 2 000 | 0.9965(10) | 0.94(3) | 0.488(1) | 1.67(6) |
| 32 | 100 000 | 2 000 | 0.9985(3) | 0.95(1) | 0.5997(9) | 4.15(11) |
| 64 | 300 000 | 2 000 | 0.99945(7) | 0.96(1) | 0.7113(9) | 14.2(4) |
| 128 | 300 000 | 2 000 | 0.99962(4) | 0.95(1) | 0.8213(18) | 58.2(3.3) |

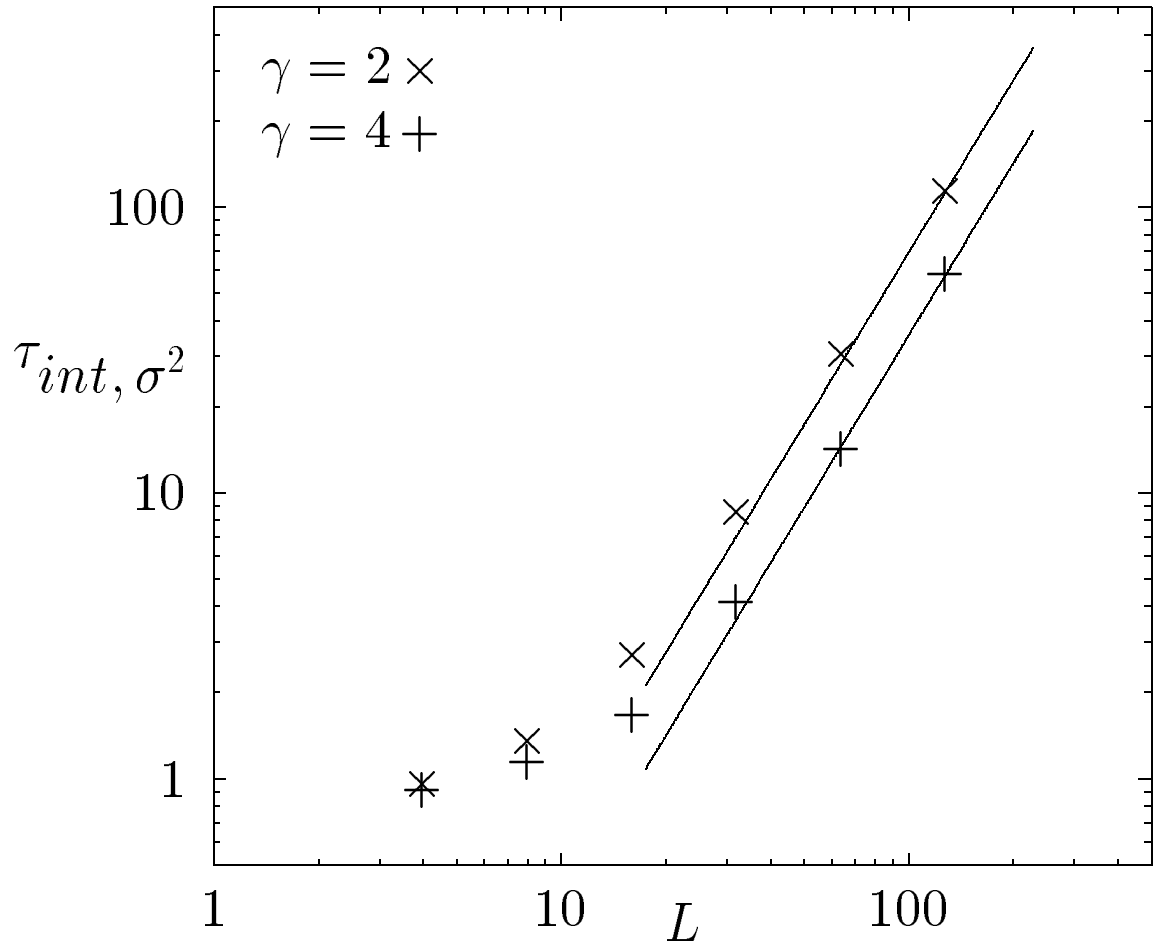

FIG. 1. Dependence of the integrated autocorrelation time for the surface thickness $\sigma^2$ on the lattice size $L$ in the rough phase of the 2-d Sine Gordon model, $\beta = 1.0$, $\zeta = 0.5$. Errors are smaller than the symbols used. The lines correspond to $z = 2$.